\documentclass[aip,jap,amsmath,amssymb,reprint,numerical,graphicx]{revtex4-1}
\usepackage{graphicx}
\usepackage{dcolumn}
\usepackage{bm}
\begin{document}
\title{Josephson Current in Ballistic Graphene Corbino Disk}
\author{Babak Abdollahipour}
\email{b-abdollahi@tabrizu.ac.ir} \affiliation{Faculty of Physics, University of Tabriz, Tabriz 51666-16471,
Iran}
\author{Ramin Mohammadkhani}
\author{Mina Khalilzadeh}
\affiliation{Department of Physics, Faculty of Science, University of Zanjan, Zanjan 45371-38791, Iran}
\begin{abstract}
We solve Dirac-Bogoliubov-De-Gennes (DBdG) equation in a superconductor-normal graphene-superconductor (SGS)
junction with Corbino disk structure to investigate the Josephson current through this junction. We find that
the critical current $I_c$ has a nonzero value at Dirac point in which the concentration of the carriers is
zero. We show this nonzero critical current depends on the system geometry and it decreases monotonically to
zero by increasing the ratio of the outer to inner radii of the Corbino disk ($R_2/R_1$), while in the limit
of $R_2/R_1\rightarrow1$ it scales like a diffusive Corbino disk. The product of the critical current and the
normal-state resistance $I_cR_N$ attains the same value for the planar structure at zero doping. These results
reveals the pseudodiffusive behavior of the graphene Corbino Josephson junction similar to the planar
structure.
\end{abstract}
\maketitle
%

%
\section{Introduction}
%
A dissipationless current in equilibrium could exist between two superconductors separated by a thin
insulating layer and its value would be proportional to the sine of the phase difference of the
superconductors order parameters, which is called the Josephson effect\cite{josephson1962}. Further studies
have shown that Josephson effect can exist if superconductors are connected by a weak link (for a review see
Ref.~\onlinecite{likharevreview1979}) in which superconducting correlations can propagate through a weak link
material via the process of retro Andreev reflection (AR) at theirs interfaces\cite{andreevspj1964}.
Conversion of the subgap electron and hole excitations with opposite spin directions to each other by
successive retro AR at two interfaces leads to the formation of a supercurrent. The Josephson effect is
characterized by the critical current $I_c$ (maximum of the Josephson current) which is characteristic of the
strength of weak link and its geometry.

The field of Josephson junctions received new attention recently, after it was recognized that suitable
Josephson devices might serve as quantum bits (qubits) in quantum information devices and that quantum logic
operations could be performed by controlling gate voltages or magnetic fields\cite{Makhlin01,Clarke08}. This
system is attractive because the low dissipation inherent to superconductors make possible, in principle, long
coherence times. In addition, because complex superconducting circuits can be microfabricated using
integrated-circuit processing techniques, scaling to a large number of qubits should be relatively
straightforward\cite{Pashkin03,Mooij05}. Further, reading out of a state of superconducting qubit has been
realized experimentally by using ballistic Josephson vortices (fluxon)\cite{Fedorov13,Fedorov14}. It is
performed by measuring the microwave radiation induced by a fluxon moving in an annular Josephson junction.
These studies and experiments motivate us to explore graphene Josephson junctions with Corbino geometry.
Results of this study may be helpful in realizing new type of superconducting qubits or in reading out of
their states.

Graphene, a two dimensional single layer of graphite which has been isolated by Novoselov \textit{et
al}\cite{novoselovscience2004}, shows a unique electronic properties due to its peculiar gapless
semiconducting band structure\cite{novoselovnature2005,geimnaturemat2007,katsnelsonsolid2007}. The conduction
and valence bands in graphene touch each other at two inequivalent corners of the hexagonal Brillouin zone in
reciprocal space, normally called $K$ and $K^{'}$ points or Dirac points. The dispersion relation of the
quasiparticle excitations about these $K$ and $K^{'}$ points is linear and obey Dirac equation (for a review
see Ref. \onlinecite{andopsj2005}). The presence of such low-lying excitations leads to unusual properties for
graphene including specular Andreev reflection at the interface with a superconductor\cite{beenakkerprl2006}
and Klein tunneling in \textit{p-n} junctions\cite{beenakkerrmp2008}. These important features of graphene has
attracted intense theoretical and experimental attention to study the effect of the relativistic-like dynamics
of electrons on Josephson effect which are already known in ordinary conducting systems.

In the last several years graphene has become a new class of weak link materials in Josephson junctions. The
graphene Josephson junction has been studied
theoretically\cite{titovprbr2006,aliapplphya2006,aliapplphya2007,Hagymasi2010,Schaffer2008,Schaffer2010,Alidoust2011,Gonzalez2008}
and experimentally\cite{Heersche2007,Miao2007,Shailos2007,Du2008,Calado2015,English2016}. Titov and Beenakker
have considered a planar structure in which two superconductors are connected to each other via an undoped
strip of ballistic graphene\cite{titovprbr2006}. Using Dirac-Bogoliubov-de Gennes
equation\cite{beenakkerprl2006}, they have shown that a nonzero supercurrent can flow through a ballistic
Josephson junction even at the Dirac point in which the carriers concentration is zero. They have found that
the critical current $I_c$ for a wide graphene junction at the Dirac point has the same form as in an ordinary
disordered normal metal\cite{beenakkerprl1991}. The Josephson effect has been studied in the graphene
nanoribbons of length $L$ smaller than superconducting coherence length and arbitrary width $W$ with smooth,
armchair and zigzag edges by Moghaddam and Zareyan\cite{aliapplphya2006,aliapplphya2007}. They have obtained
that in contrast to an ordinary superconducting quantum point contact (SQPC) the supercurrent $I_c$ in smooth
and armchair ribbons with a low concentration of the carriers is not quantized but decreases monotonically by
decreasing $W/L$. At higher concentrations of the carriers this monotonic variation acquires a series of peaks
with distances inversely proportional to the chemical potential $\mu$. The phase, the temperature and the
junction length dependence of the supercurrent for ballistic graphene Josephson-junctions have been studied by
Hagym\'{a}si \textit{et.al.}\cite{Hagymasi2010}. Black-Schaffer and Doniach have used a tight-binding
Bogoliubov-de Gennes (BdG) formalism to self-consistently calculate the proximity effect, Josephson current,
and local density of states in ballistic graphene Josephson junctions\cite{Schaffer2008}. They have shown that
self-consistency does not notably change the current-phase relationship (CPR) derived earlier for short
junctions using the non-self consistent Dirac-BdG formalism. The self-consistent temperature dependence of CPR
in ballistic graphene Josephson junctions have been studied by Black-Schaffer and Linder\cite{Schaffer2010}.
Moreover, the effect of the strain on the supercurrent in a ballistic graphene Josephson junction have been
studied by Alidoust and Linder\cite{Alidoust2011}. They have shown the supercurrent at the charge neutrality
point can be tuned efficiently by means of mechanical strain. The many-body effects on the critical current in
graphene Josephson junction has been investigated by Gonzalez and Perfetto\cite{Gonzalez2008}.

Heersche \textit{et.al.} have experimentally shown that the superconducting correlations can penetrate into
the graphene layer via the proximity effect \cite{Heersche2007}. They have confirmed the theoretical
prediction that a finite supercurrent can flow at zero charge density. Most of the experimental studies of the
graphene Josephson Junctions have been limited to the case of diffusive transport through graphene with poorly
defined and modest quality graphene-superconductor interfaces\cite{Miao2007,Shailos2007,Du2008}. Recently, a
ballistic graphene Josephson junction with a well defined and transparent interface to the graphene has been
developed using Molybdenum Rhenium contacts\cite{Calado2015}. It has been shown that the critical current
oscillates with the carrier density due to phase coherent interference of the electrons and holes that carry
the supercurrent. Further, direct measurements of the CPR for Josephson junctions with a graphene barrier has
been achieved recently by a phase-sensitive SQUID interferometry technique\cite{English2016}.

In this paper, we investigate the Josephson current in a superconductor-normal graphene-superconductor (SGS)
junction with Corbino disk structure. The SGS junction under study consists of a ballistic graphene ring
surrounded by an inner superconducting lead of radius $R_1$ and an outer superconducting lead of radius $R_2$
(Fig. \ref{fig1}). The Corbino geometry has advantage over the planar structure specially when a magnetic
field is applied to the sample due to the edge absence\cite{Hadfield03,Yan2010}. Electronic transport in
graphene Corbino structures have been studied by investigating various physical properties of these junctions
in the normal state such as conductance\cite{Rycerz09}, magnetoconductance\cite{Rycerz10},
magnetopumping\cite{Abdollahipour16} and Andreev billiards\cite{csertiprb2009}. Here, we extend these studies
to analyzing of the Josephson current in Corbino structure by solving Dirac-Bogoliubov-de Gennes (DBdG)
equation in the polar coordinate and then calculating the Andreev bound states. We show that a critical
current can flow through the junction in the limit of zero concentration of the carriers, \textit{i.e.} at
Dirac point. Further, we find that the minimal of the critical current depends on the system geometry and
reduces as the ratio of the outer to inner radii $R_2/R_1$ increases.

The paper is organized as follows. In the next section we introduce our model of a ballistic graphene
Josephsen junction with Corbino disk structure and find the solutions of the DBdG equation in the polar
coordinate. Then, we obtain the Josephson current by calculating the Andreev bound states energies. Section
III is devoted to presentation of the results and discussions. Finally, we end with a conclusion in Sec. IV.
%
\section{Model and Basic Equations} \label{sec2}
%
%
\begin{figure}
\centerline{\includegraphics[width=8cm]{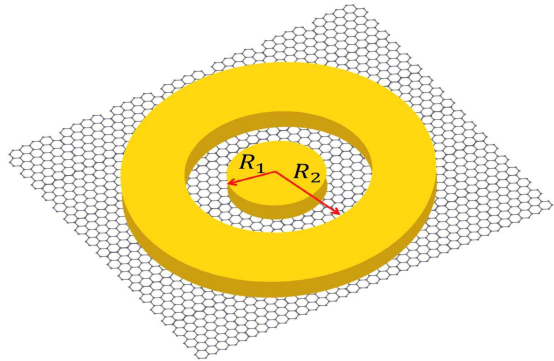}} \caption{(Color online) Schematic picture of the graphene
Corbino Josephson junction composed of two circular shaped superconducting leads connected via a ballistic
graphene ring.} \label{fig1}
\end{figure}
%
The structure under study is a graphene Corbino Josephson junction composed of a ballistic graphene ring with
the inner radius $R_1$ and the outer radius $R_2$ connected to two circularly shaped spin singlet
superconducting regions, schematically depicted in Fig.\ref{fig1}. Superconductivity in graphene is induced
via proximity effect by depositing of superconducting electrodes on top of the graphene sheet. We assume that
the superconducting regions are heavily doped, such that the Fermi wavelength inside superconducting regions
$\lambda'_{F}$ is smaller than the Fermi wavelength of graphene $\lambda_{F}$ and the superconducting
coherence length $\xi$. Thus we can neglect the suppression of the order parameter $ \Delta(\textbf{r}) $ in
the superconductors close to the interfaces and we approximate the superconducting order parameter by a step
function. Thus, it has constant values $ \Delta=\Delta_0 e^{\pm i \phi/2} $ inside superconductors and
vanishes identically in normal graphene region\cite{beenakkerprl2006}.

Low energy electron and hole excitations in graphene are described by the Dirac-Bogoliubov-De Gennes (DBdG)
equation\cite{beenakkerprl2006},
\begin{equation}\label{eq1}
\left( \begin{array}{cc} \hat{H}_{\tau} -\mu & \hat{\Delta} \\ \hat{\Delta}^{*} & \mu - \hat{H}_{\tau}
\end{array} \right) \Psi=\varepsilon \Psi\ ,
\end{equation}
where $\Psi=\left( \psi_e ~~ \psi_h \right)^T$ is the four-component wave function in the pseudospin and numbo
space, $\hat{\Delta}=\Delta\sigma_0$ is the superconducting pair potential matrix which couples time-reversal
electron and hole wave functions $\psi_e$ and $\psi_h$, $ \varepsilon > 0 $ denotes the excitation energy
measured relative to the chemical potential or the Fermi energy $\mu$. The single-particle Hamiltonian in
graphene is the two dimensional Dirac Hamiltonian in each valley $\tau=\pm$, given by
\begin{equation}\label{eq2}
\hat{H}_{\pm}=-i \hbar v(\sigma_x \partial_x \pm \sigma_y \partial_y) +V(r) ,
\end{equation}
where, $\upsilon$ denotes the Fermi velocity of the quasiparticles in graphene and $\sigma_{i=x,y,z}$ are the
Pauli matrices in the sublattice space (pseudospin) with $  \sigma_{0} $ representing the $2\times 2$ unit
matrix. In the above equation $V(r)=-U_0\Theta(-r+R_1)\Theta(r-R_2)$ denotes the electrostatic potential
throughout the system, which is zero in the normal region and nonzero in the superconductors. Since two
valleys are decouple we can solve the Hamiltonian for each valley separately, so we only consider
$\hat{H}_+=\hat{H}$.

To solve DBdG equation in a Corbino disk geometry, we write the Hamiltonian $\hat{H}$ in the polar coordinates
$(r,\varphi)$ as\cite{csertiprb2009,recherprb2007},
\begin{eqnarray}\label{eq3}
\hat{H} &=& -i \hbar \upsilon (\cos \varphi \sigma_x + \sin \varphi \sigma_y)\partial_x \nonumber \\
&&-i \hbar \upsilon (\cos \varphi \sigma_y - \sin \varphi \sigma_x)\frac{1}{r}\partial_{\varphi}+ V(r).
\end{eqnarray}
Since the Hamiltonian $ \hat{H} $ commutes with the total angular momentum $ J_z =l_z + \hbar\sigma_z/2 $,
where $l_z=-i\hbar\partial_\varphi$ is the orbital angular momentum in the $z$ direction, its eigenstates $
\psi $ are simultaneous eigenstates of $ J_z $. Thus, we can write
\begin{equation}\label{eq4}
\Psi=e^{i(m-\frac{1}{2})\varphi} \Bigg( \begin{array}{c} u_{1}(\vec{r}) \vspace*{2mm}\\
u_{2}(\vec{r})e^{i\varphi}\end{array}\Bigg)
\end{equation}
where $m=\pm1/2, \pm3/2, ... $, is a half-odd integer corresponding to the angular momentum quantum number.
For a constant electrostatic potential wave functions $ u_{1}(\vec{r}) $ and $ u_{2}(\vec{r}) $ are solutions
of the Bessel's differential equation of orders $ m-1/2$ and $m+1/2$, respectively. Therefore, in the normal
region $R_1<r<R_2$, where $V(r)=\Delta=0$, the electron and hole like eignstates of the DBdG equation with
energy $ \varepsilon $ are given by,
\begin{equation}\label{eq5}
\Psi_{N}^{e+,(-)}=e^{i(m-\frac{1}{2})\varphi}\left( \begin{array}{c} H^{(1),(2)}_{m-\frac{1}{2}}(k_e r)
\\ i \mathrm{sign}(\mu +\varepsilon )e^{i \varphi}H^{(1),(2)}_{m+\frac{1}{2}}(k_e r)\\0\\0 \end{array} \right)
\end{equation}
\begin{equation}\label{eq6}
\Psi_{N}^{h-,(+)}=e^{i(m-\frac{1}{2})\varphi}\left( \begin{array}{c} 0\\0\\H^{(1),(2)}_{m-\frac{1}{2}}(k_h
r)\\ i \mathrm{sign}(\mu - \varepsilon )e^{i \varphi}H^{(1),(2)}_{m+\frac{1}{2}}(k_h r) \end{array} \right)
\end{equation}
where $H^{(1),(2)}_{m-\frac{1}{2}}(k_{e,h} r)$ are Hankel functions of the first and second kinds and
$k_{e,(h)} = |\mu +(-) \varepsilon |/(\hbar v)$ denote wave vectors for electrons and holes. In the
superconducting leads where the pair potential is $\Delta_{L,R}$, solutions of the DBdG equation are mixed
electron-hole excitations. For the inner superconductor ($r<R_1$) the eignstates which are evanescent in this
region are given by,
\begin{equation}\label{eq7}
\Psi_{S_1}^{\pm}= e^{i (m-\frac{1}{2})\varphi} \left( \begin{array}{cc}
e^{\pm i \beta} H^{(1),(2)}_{m-\frac{1}{2}}(k_{\pm}r)\\
i e^{\pm i \beta}H^{(1),(2)}_{m+\frac{1}{2}}(k_{\pm}r) e^{i \varphi} \\
e^{-i\frac{\phi}{2}}H^{(1),(2)}_{m-\frac{1}{2}}(k_{\pm}r)\\i
e^{-i\frac{\phi}{2}}H^{(1),(2)}_{m+\frac{1}{2}}(k_{\pm}r) e^{i \varphi}
\end{array} \right)\ ,
\end{equation}
and for the outer superconductor ($r>R_2$ ) the evanescent eignstates are,
\begin{equation}\label{eq8}
\Psi_{S_2}^{\pm}= e^{i (m-\frac{1}{2})\varphi} \left( \begin{array}{cc}
e^{\mp i \beta} H^{(1),(2)}_{m-\frac{1}{2}}(k_{\mp}r)\\
i e^{\mp i \beta}H^{(1),(2)}_{m+\frac{1}{2}}(k_{\mp}r) e^{i \varphi} \\
e^{i\frac{\phi}{2}}H^{(1),(2)}_{m-\frac{1}{2}}(k_{\mp}r)\\i
e^{i\frac{\phi}{2}}H^{(1),(2)}_{m+\frac{1}{2}}(k_{\pm}r) e^{i \varphi}
\end{array} \right)
\end{equation}
where $\beta=\arccos(\epsilon/\Delta_0)$ and $ k_{\pm}=(\mu+U_0 \pm
\sqrt{\varepsilon^{2}-\Delta_{0}^{2}})/\hbar v$. The Josephson current through the graphene region is
determined by the phase difference $ \phi $ between the superconducting order parameters of two
superconductors $ S_1 $ and $ S_2 $. This supercurrent is carried by the Andreev bound states, which are
formed in the normal graphene region due to the successive conversion of the electron-hole excitations to each
other at the normal-superconductor interfaces by Andreev reflection processes. At zero temperature and in the
short-junction regime that the separation of the two NS interfaces is small with respect to the
superconducting coherence length $\xi$ ($ L \ll \xi $), the Andreev bound states (discrete spectrum) with
energies $ |\varepsilon |\leq \Delta_{0} $ have the main contribution to the supercurrent. In this limit the
Josephson current can be expressed as the following\cite{beenakkerprl1992}
\begin{equation} \label{eq2.9}
I(\phi)=-\frac{4e}{\hbar}\frac{d}{\mathrm{d} \phi} \int_0^\infty \mathrm{d}\varepsilon \sum_{n=0}^{\infty}
\rho_{n}(\varepsilon , \phi) \varepsilon\ ,
\end{equation}
where the factor 4 accounts for the spin and valley degeneracies and $\rho_{n}(\varepsilon , \phi)$ is the
density of Andreev bound states. Substitution of $\rho_{n}(\varepsilon , \phi)=
\delta[\varepsilon-\varepsilon_{n}(\phi)] $, with $\varepsilon_{n}(\phi)$ denoting the discrete spectrum of
the Andreev bound states, into Eq. (\ref{eq2.9}) gives the supercurrent as,
\begin{eqnarray}\label{eq10}
I(\phi)&=&-\frac{4e}{\hbar}\frac{d}{\mathrm{d} \phi} \int_0^\infty \mathrm{d}\varepsilon
\sum_{n=0}^{\infty} \delta[\varepsilon-\varepsilon_{n}(\phi)]  \varepsilon \nonumber \\
&&= -\frac{4e}{\hbar}  \sum_{n=0}^{\infty} \frac{d}{\mathrm{d} \phi} \varepsilon_{n}(\phi)
\end{eqnarray}
To find the energy spectrum of the Andreev bound states $ \varepsilon_n(\phi) $, we follow the approach
introduced in Ref. \cite{beenakkerprl1992}. Solutions of the wave functions inside the superconductors are
rather mixed electron-hole excitations and the interfaces scatter the particles between the two neighboring
regions. However, the excited quasiparticles located in the N layer cannot penetrate directly into a
superconductor if its energy is smaller than the superconducting pair potential and consequently two kinds of
processes, Andreev and normal reflection can occur for them. A simple mode matching at the NS interfaces
(namely, $r=R_1$ and $r=R_2$) gives the result for Andreev scattering matrix for the conversion from electron
to hole $\hat{r}_{he}$ and $\hat{r}_{eh}$ for that from hole to electron
\begin{eqnarray}\label{eq11}
\hat{r}_{he}=e^{-i\beta}\left( \begin{array}{cc} e^{-i\varphi_L} & 0 \\ 0 & e^{-i\varphi_R}
\end{array} \right)\ ,
\\\nonumber
\hat{r}_{eh}=e^{-i\beta}\left( \begin{array}{cc} e^{i\varphi_L} & 0 \\ 0 & e^{i\varphi_R}
\end{array} \right)\ .
\end{eqnarray}
In obtaining these results we have considered that the Fermi wave length $\lambda^{\prime}_F=\hbar
v/(E_F+U_0)$ in S is sufficiently small than the Fermi wave length $\lambda_F=\hbar v/E_F$ in N and the
superconducting coherence length $\xi=\hbar v/\Delta_0$. This condition is satisfied simply by taking the
limit of $U_0\rightarrow\infty$. In the normal region the quantum transport of the electron (hole) is
described by the scattering matrix $\hat{S}_e$($\hat{S}_h$). The scattering matrices are related to each other
by $\hat{S}_h(\epsilon)=\hat{S}_e(-\epsilon)^{*}$. A round trip for the electron or hole wave functions
results in a equation for product of the scattering matrices
\begin{equation} \label{det}
\textit{Det}\left(\hat{1}-\hat{r}_{eh}\hat{S}_{h}\hat{r}_{eh}\hat{S}_{e}\right)=0\ ,
\end{equation}
We can simplify this equation in the short-junction limit $L\ll\xi$ that the length $L$ of the normal region
is small relative to the superconducting coherence length, which is experimentally most relevant for
superconductors with small gap. In terms of the energy scales this condition is equivalent to
$\Delta_0\ll\hbar/\tau_{dwell}=\hbar v/L$, where $\tau_{dwell}$ is the dwell time in the junction. In this
regime we may approximate
$\hat{S}_{e(h)}(\epsilon)\simeq\hat{S}_{e(h)}(-\epsilon)\simeq\hat{S}_{e(h)}(0)\equiv\hat{S}^{(*)}_0$ for
$\epsilon$ of order $\Delta_0$. Using the exact form of the scattering matrix,
\begin{eqnarray}\label{eq13}
\hat{S}_{0}=\left( \begin{array}{cc} r_{11} & t_{12} \\ t_{21} & r_{22}
\end{array} \right)\ ,
\end{eqnarray}
and its unitarity $\hat{S}^{*}_0\hat{S}_0=\hat{1}$, we can reduce the Eq. \ref{det} to the simpler form of
\begin{equation} \label{eq14}
\textit{Det}\left[(1-\epsilon^2/\Delta^2_0)\hat{1}-t_{12}t^{\dagger}_{12}\sin^2(\phi/2)\right]=0\ ,
\end{equation}
where $\phi=\phi_R-\phi_L$ is the phase difference between superconductors. We can solve this equation for
$\epsilon$ in terms of the eigenvalues of $T=t_{12}t^{\dagger}_{12}$. Finally, energies of the Andreev bound
states are given by\cite{beenakkerprl1992},
\begin{equation} \label{eq15}
\epsilon_{n}=\Delta_0\sqrt{1-T_{n}\sin^2(\phi/2)}~~~,~ n=0,1,2,...
\end{equation}
Using this result Eq. \ref{eq11} for Josephson current is reduced to
\begin{eqnarray}\label{eq16}
I(\phi)=\frac{e\Delta_0}{\hbar} \sum_{n=0}^{\infty} \frac{T_n\sin\phi}{\sqrt{1-T_n\sin^2(\phi/2)}}
\varepsilon_{n}(\phi) \ .
\end{eqnarray}
Further, the normal state resistance $R_N$ is given by
\begin{eqnarray}\label{eq17}
1/R_N=\frac{4e^2}{\hbar} \sum_{n=0}^{\infty} T_n \ .
\end{eqnarray}
The normal state transmission probability $T_n$ is calculated by a simple mode matching in the normal state
structure\cite{Rycerz09}. Using the wave function given by Eq. \ref{eq5} for normal state and taking the limit
of $U_0\rightarrow\infty$ we can obtain the transmission probabilities as\cite{Rycerz09},
\begin{eqnarray}\label{eq18}
T_n=\frac{16}{\pi^2(kR_1)(kR_2)}\frac{1}{(\Gamma^+_n)^2+(\Gamma^-_n)^2} \ ,
\end{eqnarray}
where
\begin{eqnarray}\label{eq19}
\Gamma^{+(-)}_n=\textit{Im}\left[H^{(1)}_{n}(kR_1)H^{(2)}_{n(n+1)}(kR_2)\right.\nonumber\\\left.+(-)
H^{(1)}_{n+1}(kR_1)H^{(2)}_{n+1(n)}(kR_2)\right] \ ,
\end{eqnarray}
with $k=\mu/\hbar v$ and $n=m-1/2$. Now we have all requirements for calculating the Josephson current in
Corbino Josephson junction. In the next section we will give the results for critical current in this
junction.

\section{Results and Discussion\label{sec3}}

%
\begin{figure}
\centerline{\includegraphics[width=7cm]{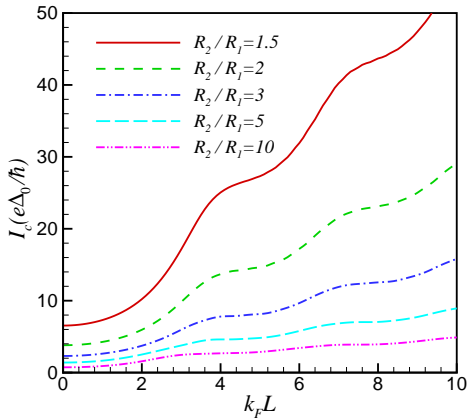}} \caption{(Color online) Critical Josephson current $I_c$ of
a Corbino Josephson junction as a function of the reduced chemical potential $k_FL$ for different values of
$R_2/R_1$.} \label{fig2}
\end{figure}
%
%
\begin{figure}
\centerline{\includegraphics[width=7cm]{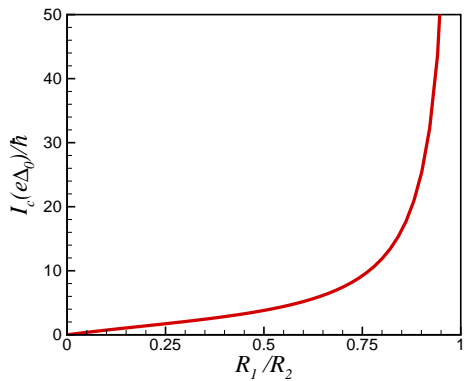}} \caption{(Color online) Critical Josephson current $I_c$ of
a Corbino Josephson junction as a function of radii aspect ratio of Corbino disk $R_1/R_2$ at zero doping.}
\label{fig3}
\end{figure}
%

\begin{figure}
\centerline{\includegraphics[width=7cm]{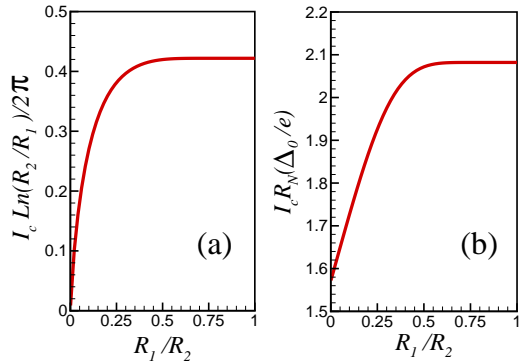}} \caption{(Color online) (a) Normalized critical Josephson
current $I_c\textit{Ln}(R_2/R_1)/2\pi$ and (b) product of the critical current and normal-state resistance as
a function of radii aspect ratio of Corbino disk $R_1/R_2$ at zero doping.} \label{fig4}
\end{figure}

In this section, we present results for the critical current $I_c$ in terms of the reduced chemical potential
$k_FL=\mu L/\hbar v$, with $L=R_2-R_1$, and ratios of the outer to inner radii of Corbino disk $R_2/R_1$, at
zero temperature. Fig. \ref{fig2} shows the critical current in terms of the reduced chemical potential for
different $R_2/R_1$. Similar to the planar junction, critical current has an oscillatory behavior and
increases by increasing $k_FL$. This oscillatory behavior is due to phase coherent interference of the
electrons and holes that carry the supercurrent caused by the formation of a Fabry-P\'{e}rot cavity.
Increasing $R_2/R_1$ results in decreasing of $I_c$ in response to the junction length ($L=R_2-R_1$) increment
and critical current goes to have a linear dependence on $k_FL$ at high values of radii ratio $R_2/R_1$. The
graphene Corbino Josephson junction supports a nonzero minimal supercurrent at the zero doping identical to
the planar graphene Josephson junctions. This behavior is caused by the evanescent modes which are exist at
the Dirac point in combination with the Klein tunneling effect. We have plotted the critical current in terms
of $R_1/R_2$ in Fig. \ref{fig3} to compare minimal supercurrent at zero doping for different values of the
radii aspect ratio. This figure shows the scaling behavior of the critical current at zero doping. To analyze
this nonzero minimal supercurrent we notice that at zero doping and in the limit of $R_2/R_1\simeq 1$, the
normal state conductivity of the Corbino junction scales by the factor
$2\pi/\textit{Ln}(R_2/R_1)$\cite{Rycerz09}. In the opposite limit $R_2\gg R_1$, conductance of the normal
junction scales by $8\pi(R_1/R_2)$. Thus, as it is apparent in Fig. \ref{fig3}(a), the critical current takes
a Logarithmic scaling when it approaches $R_1/R_2=1$, and it vanishes by increasing $R_2/R_1$ and tends
linearly to zero in the limit of $R_1/R_2\rightarrow 0$. The normalized critical Josephson current,
$I_c\textit{Ln}(R_2/R_1)/2\pi$, takes a constant value for $R_1/R_2>0.5$ as it has been shown in Fig.
\ref{fig4}(a). Further, as  Fig. \ref{fig4}(b) shows, the product of the critical Josephson current and
normal-state resistance $I_cR_N$, approaches to its limiting value at $R_1/R_2\rightarrow 1$ for
$R_1/R_2>0.5$, and takes a nonzero value ($\sim 1.57$) at the limit of $R_1/R_2\rightarrow 0$. Since $R_N$ has
inverse scaling with respect to $I_c$, their product $I_cR_N$ takes nonzero value for all values of the
$R_1/R_2$ at zero doping.
%
%

The limiting behavior of the critical current for $R_2/R_1\rightarrow 1$ at zero doping can be deduced by
investigating the limiting behavior of the transmission probability given by Eq. \ref{eq18}. At zero doping
the transmission probability is given by \cite{Rycerz09},
\begin{equation} \label{Iphi}
T_j=\frac{1}{\cosh^2[j\textit{ln}(R_1/R_2)]}\ ,~~~~j=\frac{1}{2},\frac{3}{2},\frac{5}{2},...
\end{equation}
Then using Eq. \ref{eq10} we can obtain the following expression for the Josephson current at zero doping and
in the limit of $R_2/R_1\rightarrow 1$,
\begin{eqnarray} \label{Iphi}&&
I(\phi)=\frac{e\Delta_0}{\hbar}\frac{4}{Ln(R_2/R_1)}\cos(\phi/2)\textit{arctanh}[\sin(\phi/2)]\ ,
\nonumber\\&& I_c=1.33\frac{e\Delta_0}{\hbar}\frac{2}{Ln(R_2/R_1)}\ , \nonumber\\&&
I_cR_N=2.08\frac{\Delta_0}{e}. \label{limit1}\end{eqnarray}
These expressions match with the results presented in Figs. \ref{fig3} and \ref{fig4}(a)(b). These results for
ballistic graphene with Corbino geometry at the Dirac point are similar to those of a planar junction of a
ballistic graphene or a disorder normal metal. Therefore, we revisit the pseudodiffusive behavior at zero
doping for the graphene Corbino Josephson junction\cite{titovprbr2006}.

\begin{figure}
\centerline{\includegraphics[width=7cm]{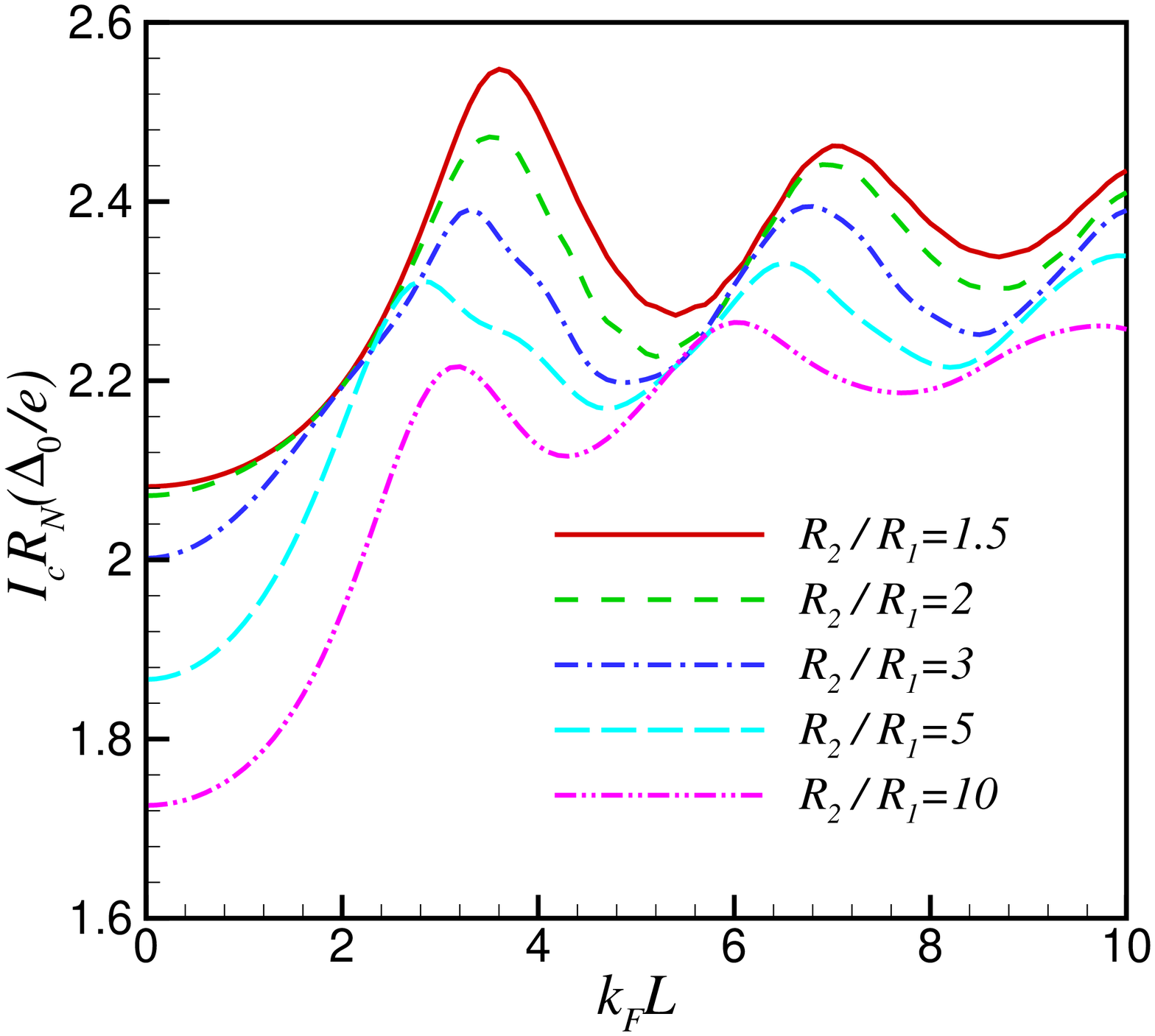}} \caption{(Color online) Product of the critical Josephson
current and normal-state resistance $I_cR_N$ of a Corbino Josephson junction as a function of the reduced
chemical potential $k_FL$ for different values of $R_2/R_1$.} \label{fig5}
\end{figure}

To complete our discussion, the dependence of $I_cR_N$ on $k_FL$ for different values of $R_2/R_1$ has been
depicted in Fig. \ref{fig5}. As these results show $I_cR_N$ reduces by increasing $R_2/R_1$ for all values of
the carrier concentration. At the zero doping and in the limit of $R_2/R_1\rightarrow 1$, $I_cR_N$ is
correctly given by Eq. \ref{limit1}. On the other hand, at the limit of $k_FL\gg 1$ it approaches to the value
given by $I_cR_N=2.44\Delta_0/e$ identical to the planar graphene Josephson junctions. Recent realization of
the ballistic graphene Josephson junctions with well defined superconductor and graphene
contacts\cite{Calado2015} and the resistance measurements performed on the graphene Corbino
junction\cite{Yan2010} makes it possible to observe the scaling behavior of the graphene Corbino Josephson
junction reported here.
%
%

\section{Conclusions \label{sec4}}

In conclusion we have studied the Josephson current in the graphene Corbino Josephson junction. The Corbino
Josephson junction has composed of a ring shape graphene attached to two coaxial superconducting leads. We
have analyzed the Josephson current in the short junction limit by calculating the Andreev bound state
energies. We have shown that the ballistic Corbino Josephson junction supports a nonzero critical Josephson
current at the zero doping and behave similar to the diffusive Corbino Josephson junctions in the zero doping
limit. This result revealed the pseudodiffusive behavior of the Corbino Josephson junction at zero doping
identical to the planar junction. The critical Josephson current decreases monotonically by increasing the
outer to inner radii ratio of Corbino junction. Moreover, we have shown that the product of the critical
current and the normal-state resistance have a nonzero value for all values of the aspect ratios of radii of
Corbino disk. The scaling behavior studied here can be observed in the experiment.


\end{document}